\documentclass[%
reprint,
 amsmath,amssymb,
 aps,
]{revtex4-1}

\usepackage{graphicx}
\usepackage{dcolumn}
\usepackage{bm}

\begin{document}

\preprint{APS/123-QED}

\title{Superconducting gap symmetry of the BiS$_2$-based superconductor LaO$_{0.5}$F$_{0.5}$BiSSe elucidated through specific heat measurement}

\author{Naoki Kase$^{1}$}
 \altaffiliation{Present address: Department of Applied Physics, Tokyo University of Science, Tokyo 125-8585, Japan}
 \thanks{n-kase@rs.tus.ac.jp}
\author{Yusuke Terui$^{1}$}%
\author{Tomohito Nakano$^{1}$}
\author{Naoya Takeda$^{2}$}
\affiliation{%
 $^{1}$Graduate School of Science and Technology, Niigata University, Niigata 950-2181, Japan\\
 $^{2}$Department of Materials Science, Niigata University, Niigata 950-2181, Japan
}%




\date{\today}

\begin{abstract}
In this study, the superconducting gap symmetry of LaO$_{0.5}$F$_{0.5}$BiSSe is elucidated through specific heat measurements with one single crystal of LaO$_{0.5}$F$_{0.5}$BiSSe.
A clear jump is observed at 3.95 K in LaO$_{0.5}$F$_{0.5}$BiSSe, suggesting bulk superconductivity.
From specific heat measurements, $\Delta C_\mathrm{e}/\gamma T_\mathrm{c}$ and 2$\Delta$(0)/$k_\mathrm{B}T_\mathrm{c}$ are obtained as 2.31 and 4.5.
These values indicate strong electron--phonon coupling superconductivity.
The temperature ($T$) and magnetic-field ($H$) dependence of the electronic specific heat $C_\mathrm{e}(T, H)$ suggests that superconductivity is fully gapped.
Our handmade system successfully elucidates superconducting gap symmetry of the BiS$_{2}$-based compound with a small single crystal.
\end{abstract}

\pacs{Valid PACS appear here}
\maketitle



\section{Introduction}
Superconductivity of the BiS$_2$-based layered compounds Bi$_4$O$_3$S$_4$ and LaO$_{0.5}$F$_{0.5}$BiS$_2$ has attracted much attention\cite{1,2,3,4,5,6}. 
Their crystal structure consists of the LaO-blocking and BiS$_2$-conducting layers, and is very similar to FeAs-based high-$T_\mathrm{c}$ superconductor.
By analogy with the FeAs-based superconductor and the BiS$_2$-based superconductor, it is highly possible that the origin of the superconducting (SC) transition mechanism is unconventional.

The band structure calculation suggests the existence of a strong nesting of the Fermi surface at wave vector $\mathbf{ k}$ = ($\pi$, $\pi$)\cite{7,8, 9,10}.
The importance of the nesting is experimentally pointed out and shows the possibility that electronic correlations play an important role in the emergence of superconductivity of these systems.
However, it was argued that effect of the electron correlation is a trivial factor in this system.
The conduction band in the 2D layer of BiS$_2$-based compounds is mainly contributed from 6$p_x$ and 6$p_y$ orbits of Bi\cite{7}.
Because of the widely spread 6$p$ orbitals, the electron correlation effects appear to be unimportant.
A strong electron correlation is significant for unconventional superconductivity.
Thus, it is central issue to elucidate whether interactions that mediate the Cooper pair of superconductivity is unconventional.

Extensive theoretical studies have proposed the SC gap structures, which are intimately related to the paring interaction\cite{7,8,9,10,a1,a2,a3}.
Experimentally, unconventional superconductivity can be expected to be generated because an extremely large ratio of 2$\Delta/k_\mathrm{B}T_\mathrm{c}$ $\sim$ 17, which is about five times larger than the BCS theory is found in scanning-tunneling microscopy (STM) measurements\cite{delta}.

To determine SC gap symmetry, $\mu$SR measurements of polycrystalline Bi$_{4}$O$_{4}$S$_{3}$ and LaO$_{0.5}$F$_{0.5}$BiS$_{2}$ are performed, and suggest that superconductivity is fully gapped\cite{muon}. 
Furthermore, several measurements of a single crystal of Nd(O, F)BiS$_{2}$ imply fully gapped superconductivity. 
Raman scattering suggests a possible phonon-mediated superconductivity\cite{Raman}.
Penetration depth and thermal conductivity measurements have been reported for a fully gapped superconductor\cite{Pen, thermal}. 
However, recent angle-resolved photoemission spectroscopy (ARPES) measurement suggests that NdO$_{0.71}$F$_{0.29}$BiS$_{2}$ is an unconventional superconductor that has competitive or cooperative multiple pairing interactions\cite{arpes}.
Thus, SC symmetry of the BiS$_{2}$-based compounds is controversial.

Specific heat is one of the most significant tool for understanding SC state and to confirm bulk nature of superconductivity thermodynamically.
However, specific heat measurements have not been performed to clarify SC gap symmetry in BiS$_2$-superconductors, even though confirmation about bulk nature of superconductivity has been reported\cite{15,16,17}.
The difficulty in measuring the specific heat of BiS$_2$-compounds at low temperatures is due to the need of treating a small single crystal with a small Sommerfeld coefficient. 
To overcome the present situation, we constructed a hand-made sensitive calorimeter, which allows us to implement high precision measurements, and is a powerful tool to determine SC symmetry.

In this article, we report temperature and magnetic field dependences of specific heat $C(T, H)$ of LaO$_{0.5}$F$_{0.5}$BiSSe by using the sensitive calorimeter with one piece of single crystal (less than 1 mg).
The Se-substituted compounds LaO$_{0.5}$F$_{0.5}$Bi(S$_{1-x}$Se$_x$)$_2$ are known to emergence bulk superconductivity with the optimum Se content $x$ of 0.5\cite{11,12}.
Through our measurements, we have succeeded to elucidate the SC-gap symmetry of the BiS$_{2}$-based compound by using one piece of single crystal.


\section{Experimental Details and Results}
\begin{figure}
\begin{center}
\includegraphics[width=3.250in]{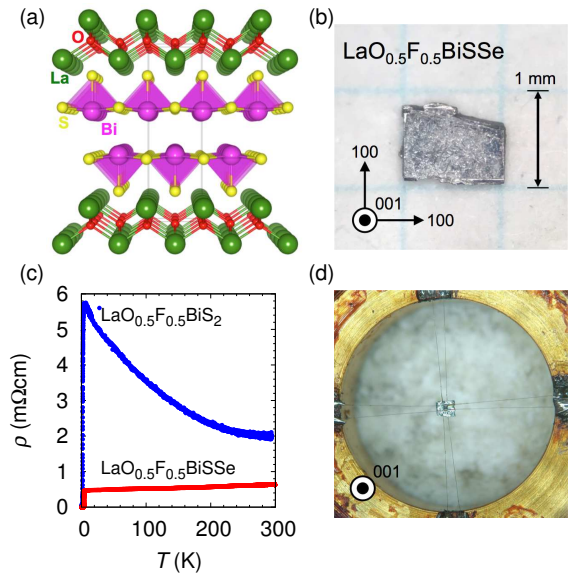}
\caption{(Color Online)
(a) Crystal structure of the BiS$_{2}$-based compounds.
(b) Single crystal of the LaO$_{0.5}$F$_{0.5}$BiSSe (0.42 mg).
(c) $T$-dependence of the electrical resistivity $\rho(T)$ of LaO$_{0.5}$F$_{0.5}$BiS$_2$ and LaO$_{0.5}$F$_{0.5}$BiSSe down to 0.5 K.
(d) Setup for specific heat measurements using our hand-made calorimeter.
}
\label{fig1}
\end{center}
\end{figure}

Single crystals of LaO$_{0.5}$F$_{0.5}$BiS$_2$ and LaO$_{0.5}$F$_{0.5}$BiSSe were grown by using a CsCl/NaCl-flux method\cite{13,14}.
As shown in Fig. \ref{fig1}(b), the typical size of the single crystals is approximately 1 $\times$ 1 mm$^{2}$.
The crystal structure of the synthesized sigle-crystalline sample was examined through powder X-ray diffraction (XRD) implemented using a conventional X-ray spectrometer equipped with Cu-K$\alpha$ radiation and a graphite monochromator (RAD-2X, Rigaku). 
The lattice constants of LaO$_{0.5}$F$_{0.5}$BiSSe were estimated to be 0.41074 and 1.3486 nm, and of LaO$_{0.5}$F$_{0.5}$BiS$_2$ were 0.40512 and 1.3410 nm.
These values are almost consistent with those of the previous reports\cite{11,12}.

\begin{figure}
\begin{center}
\includegraphics[width=3.25in]{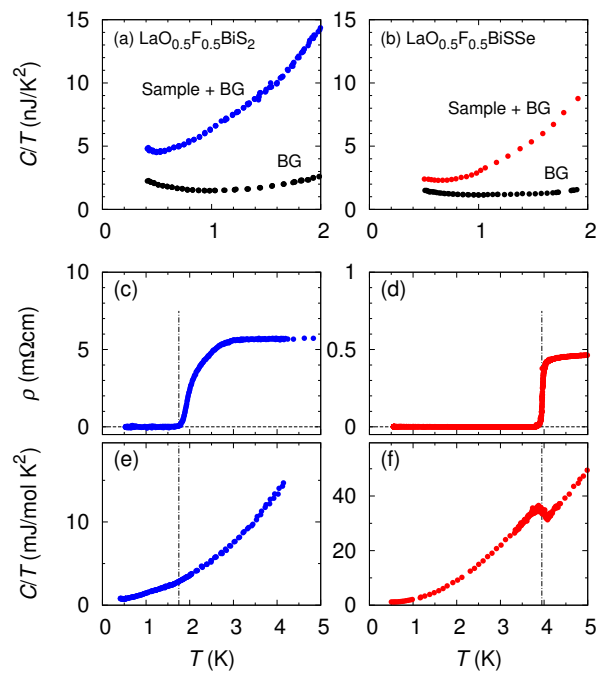}
\caption{(Color Online) 
$T$-dependence of the specific heat $C(T)$ divided by $T$ of background and raw data including sample and background of (a) LaO$_{0.5}$F$_{0.5}$BiS$_2$ and (b) LaO$_{0.5}$F$_{0.5}$BiSSe in zero magnetic field.
$T$-dependence of the electrical resistivity $\rho(T)$ of (c) LaO$_{0.5}$F$_{0.5}$BiS$_2$ and (d) LaO$_{0.5}$F$_{0.5}$BiSSe in zero magnetic field.
$T$-dependence of the specific heat $C(T)$ divided by $T$ of (e) LaO$_{0.5}$F$_{0.5}$BiS$_2$ and (f) LaO$_{0.5}$F$_{0.5}$BiSSe in zero magnetic field.
}
\label{fig2}
\end{center}
\end{figure}

Superconductivity was confirmed by the $T$-dependence of the electrical resistivity $\rho(T)$ measurements, which were performed using a conventional four-probe method with a current source (Model 6221, Keithley) and a nano voltmeter (Model 2182A, Keithley), down to 0.5 K in zero and various applied magnetic fields.
Gold wires were attached to the sample surface by Ag paste.
Figure \ref{fig1}(c) shows $\rho(T)$ of LaO$_{0.5}$F$_{0.5}$BiS$_2$ and LaO$_{0.5}$F$_{0.5}$BiSSe up to room temperature.
$\rho(T)$ of LaO$_{0.5}$F$_{0.5}$BiS$_2$ shows semi-metallic behavior with a residual resistivity $\rho_0$ of approximately 5.8 m$\Omega$cm: whereas $\rho(T)$ of LaO$_{0.5}$F$_{0.5}$BiSSe behaves as almost $T$-linear dependent with $\rho_0$ = 0.47 m$\Omega$cm.
Both compounds show superconductivity.
$T_\mathrm{c}$ in $\rho(T)$ is defined as a 50\% drop from $\rho_0$, and the superconducting transition width $\Delta T$ is taken as the temperature interval between 10\% and 90\% of $\rho_0$.
$T_\mathrm{c}$ is determined to be 2.2 and 3.95 K with $\Delta T$ = 0.7 and 0.02 K, respectively.

Only one piece of specimen is sufficient to perform specific heat measurements, because the background value is comparative to the raw data of the sample even at 0.5 K.
Specific heat measurements were performed through a relaxation method down to 0.4 K by using a commercial $^3$He refrigerator (Heliox-VL, Oxford Instruments) equipped with a superconducting magnet (Oxford instruments).
Our hand-made system consists of a heater and thermometer (RuO$_2$ chip resistor) without a sample stage.
The sample was sandwiched between the heater and thermometer, which were hung by using NbTi SC wires attached using an Ag paste on the surfaces of the RuO$_2$ chip resistor to form an electrode.
As the bottom of the chip resistor consists of an Al$_2$O$_3$ insulator, the sample cannot be electrically contacted with the heater and thermometer.
Figure \ref{fig1}(d) presents the setup for the measurement of $C(T)$.
A single crystal of LaO$_{0.5}$F$_{0.5}$BiS$_2$ (1.40 mg) and LaO$_{0.5}$F$_{0.5}$BiSSe (0.42 mg) fixed using N grease between the heater and thermometer was used for the measurements.
Figures \ref{fig2}(a) and (b) show the $T$-dependence of the specific heat $C(T)$ divided by $T$ of raw data (sample + background) and background down to 0.4 K.
The contribution of the background is relatively smaller than that of one specimen.
The background value is comparable to that of the sample even at 0.4 K.
Thus, we determined the validity that our hand-made system enables to measure the accurate values of small single crystals of LaO$_{0.5}$F$_{0.5}$BiSSe.

\begin{figure}
\begin{center}
\includegraphics[width=3.25in]{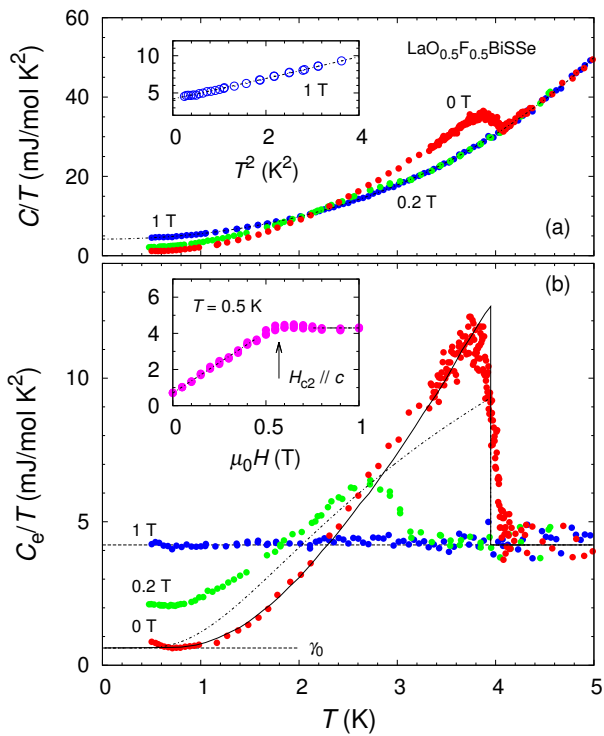}
\caption{(Color online) 
(a) $T$-dependence of the specific heat $C(T)$ divided by $T$ of LaO$_{0.5}$F$_{0.5}$BiSSe in magnetic fields of 0, 0.2, and 1 T.
The dotted line is the fitting result of $C/T$ = $\gamma + \beta T^2 + \delta T^4$.
The inset shows the $T^2$ dependence of $C(T)/T$ at the low temperature region.
(b) Electronic specific heat $C_\mathrm{e}(T)$ divided by $T$.
The solid line shows the theoretical calculation of the $\alpha$ model.
The dotted lines represent the calculated BCS curve assuming a weak coupling limit.
The inset shows the $H$-dependence of the electronic specific heat $C_\mathrm{e}(H)$ divided by $T$ at 0.5 K.
}
\label{fig3}
\end{center}
\end{figure}

Figure \ref{fig2}(e) and (f) show $C(T)/T$ of LaO$_{0.5}$F$_{0.5}$BiS$_2$ and LaO$_{0.5}$F$_{0.5}$BiSSe in zero magnetic field.
The bulk nature of the superconductivity can be confirmed because of the clear jump at $T_\mathrm{c}$ = 3.95 K, which is consistent with $T_\mathrm{c}$ determined by $\rho(T)$.
The transition is very sharp with a transition width of 0.2 K, indicating good sample quality.
However, no jump was observed in LaO$_{0.5}$F$_{0.5}$BiS$_2$, although the slope seems to change at the temperature in which $\rho(T)$ shows zero resistivity.
This result suggests that superconductivity of LaO$_{0.5}$F$_{0.5}$BiS$_2$ is not of bulk nature.
These results correspond to those of magnetic susceptibility measurements.

Figure \ref{fig3}(a) shows $C(T)/T$ of LaO$_{0.5}$F$_{0.5}$BiSSe in magnetic fields of 0 and 1 T.
To describe the specific heat data at $\mu_0 H$ = 1.0 T, which exceeds $H_\mathrm{c2}$(0), we use the following expression: $C(T)/T = \gamma + \beta T^2 + \delta T^4$, where $\gamma$ is the electronic specific heat coefficient (Sommerfeld constant), and  $\beta T^2 + \delta T^4$ (= $C_{\mathrm{ph}}$) is the lattice vibration contribution.
We obtained parameters $\gamma$ = 4.19 mJ/(mol$\cdot$K$^2$), $\beta$ = 1.34 mJ/(mol$\cdot$K$^4$), and $\delta$ = 0.0198 mJ/(mol$\cdot$K$^6$).
In a simple Debye model of the phonon contribution, the $\beta$ coefficient is related to the Debye temperature $\Theta_\mathrm{D}$ = $(\frac{12\pi^4}{5\beta} nR)^{1/3}$, where $R$ = 8.314 J/(mol$\cdot$K) is the gas constant and $n$ = 5 is the number of atoms per formula unit.
From this relationship, $\Theta_\mathrm{D}$ is estimated to be 194 K.

Next, we discuss the detailed $T$-dependence of the electronic specific heat $C_\mathrm{e}(T)$ of LaO$_{0.5}$F$_{0.5}$BiSSe.
Figure \ref{fig3}(b) shows $C_\mathrm{e}(T)$ divided by $T$ of the Se-doped sample.
$C_\mathrm{e}(T)/T$ was fitted to the theoretical curve assuming an isotropically gapped SC state, based on the following equations \cite{gap}:
\begin{equation}
\label{ }
\frac{S}{\gamma_nT_\mathrm{c}} = -\frac{6}{\pi^2}\frac{\Delta(0)}{k_\mathrm{B}T_\mathrm{c}}\int_0^\infty \left[ f\ln f + (1 - f)\ln(1 - f) \right]dy,
\end{equation}
\begin{equation}
\label{ }
\frac{C}{\gamma_nT_\mathrm{c}} = t\frac{d(S/\gamma_n T_\mathrm{c})}{dt},
\end{equation}
where $f = [\exp(\beta E) + 1]^{-1}$ and $\beta$ = $(k_\mathrm{B}T_\mathrm{c})^{-1}$.
The energy of quasiparticles is given by $E$ = $\left[ \varepsilon^2 + \Delta^2(t) \right]^{1/2}$, where $\varepsilon$ is the energy of the normal electrons relative to the Fermi energy.
In addition, the $T$-dependence of an energy gap varies as $\Delta(t)$ = $\Delta(0)\delta(t)$, where $\delta(t)$ is the normalized BCS gap, as calculated by Muhlschlegel\cite{bcs}.
The result reproduces the experimental data quite well in a wide temperature range just below $T_\mathrm{c}$ down to 0.5 K.
The entropy balance is satisfied, and we obtained $T_\mathrm{c}$ = 3.95 K and $\Delta(0)/k_\mathrm{B}T_\mathrm{c}$ = 2.25, where $\Delta$(0) is the amplitude of the SC gap at $T$ $\rightarrow$ 0 K (solid line).
The value is clearly smaller than 2$\Delta/k_\mathrm{B}T_\mathrm{c}$ $\sim$ 17 obtained from STM\cite{delta}.
For comparison, we also plot the BCS curve for the weak electron--phonon coupling limit (dotted line).

As shown in Fig. \ref{fig3}(b), a clear residual linear term $\gamma_0$ = 0.60 mJ/(mol$\cdot$K$^2$) is observed in the $T$ $\rightarrow$ 0 K limit from $C(T)$/$T$.
This may have originated from a small fraction of the non--SC phases.
After subtracting the impurity concentration $\gamma_0$, we found the Sommerfeld coefficient to be $\gamma_\mathrm{n}$ = $\gamma$ -- $\gamma_0$ = 3.59 mJ/(mol$\cdot$K$^2$).
The normalized specific heat jump was estimated to be $\Delta C_\mathrm{e}/\gamma T_\mathrm{c}$ = 2.31, which is larger than 1.43 in the weak-coupling limit.
Our results strongly suggest a strong electron--phonon coupling.

\begin{figure}
\begin{center}
\includegraphics[width=3.0in]{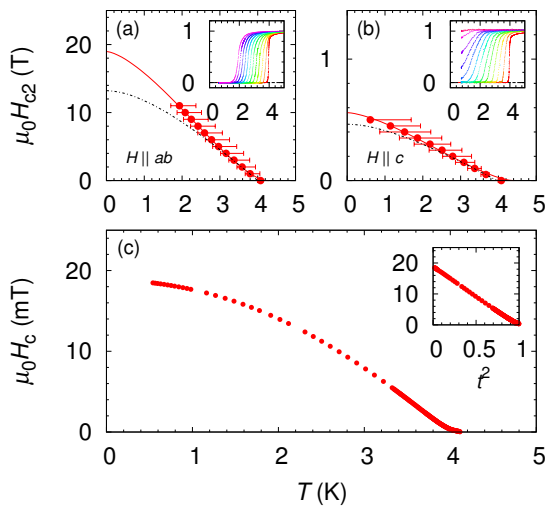}
\caption{(Color Online) 
$T$-dependence of the upper critical field $H_\mathrm{c2}$($T$) of LaO$_{0.5}$F$_{0.5}$BiSSe for $H$ $||$ $ab$ (a) and $H$ $||$ $c$ (b).
The inset shows the $T$-dependence of the electrical resistivity $\rho(T)$ normalized at $\rho_{0}$ just above $T_{\mathrm{c}}$ in several magnetic fields.
The dotted line is the WHH theory with the dirty limit ($h^\ast$ = 0.69).
(c) $T$-dependence of the thermodynamic critical field $H_\mathrm{c}$($T$).
The inset shows $H_\mathrm{c}$ vs $t^{2}$, $t$ = $T/T_\mathrm{c}$.
}
\label{fig4}
\end{center}
\end{figure}

Next, we discuss the $H$-dependence of the low-temperature part of the electronic specific heat $C_\mathrm{e}(H)$, which is known to be very sensitive to the SC structure.
The inset of Fig. \ref{fig3}(b) shows $C_\mathrm{e}(H)$ divided by $T$ at 0.5 K in magnetic field applied perpendicular to the $ab$-plane ($H || c$).
$C_\mathrm{e}(H)/T$ shows a $H$-linear dependence in a wide magnetic-field range.
Rapid increase detected from the $H$-dependence of the thermal conductivity $\kappa(H)$ of NdO$_{0.71}$F$_{0.29}$BiS$_{2}$ is not seen in $C_{e}(H)$.
We note a small upturn of $C_\mathrm{e}(T)/T$ at approximately 0.5 K in zero magnetic field.
The origin is likely to be a nuclear specific heat but
the upturn is suppressed in higher magnetic filed.
We consider that the upturn is ascribed to the background contribution, which is insufficiently subtracted from raw data\cite{0.3}.
The increase is about 0.18 and 0.015 mJ/(mol$\cdot$K$^{2}$) in magnetic fields of 0 and 0.2 T.
The effect of the upturn on $C_\mathrm{e}(H)$ is vanishingly small enough to discuss the SC gap even in 0.5 K.
Thus, $C_\mathrm{e}(H)$ also provides supporting evidence for a fully gapped SC state.

The upper critical field $\mu_0 H_\mathrm{c2}(T)$ was obtained from the $\rho(T)$ measurements in several magnetic fields, as shown in Fig. \ref{fig4}(a) and (b).
The $\mu_0 H_\mathrm{c2}(T)$ curves show clear enhancement from the Wertharmer--Helfand--Hohenberg (WHH) theory\cite{hc2b, hc2a} in the dirty limit (dotted line) and upper curvature around $T_\mathrm{c}$.
The features are usually understood as a multi-gap SC, but are also originated from spin-orbit and/or strong electron-phonon coupling\cite{hc2a,hc2}.
Because our measurements suggest single SC gap, it is inadequate to employ the multi-gap fit.
The $H_\mathrm{c2}$(0) was estimated from the fitting formula $H_\mathrm{c2}(T) = H_\mathrm{c2}(0)\left[1 - (T/T_\mathrm{c})^{3/2}\right]^{3/2}$ for superconductors with upper curvature\cite{hc2fit}. 
The $\mu_0 H_\mathrm{c2}$(0) was evaluated to be approximately 18.5 T for $H$ $||$ $ab$ and 0.572 T for $H$ $||$ $c$.
This value is consistent with that of the previous report\cite{13}.
The anisotropic parameter $\Gamma$ was determined using the anisotropic Ginzburg--Landau (GL) formula.
\begin{equation}
\label{ }
\Gamma = \frac{m_c}{m_{ab}} = \frac{H^{|| ab}_\mathrm{c2}}{H^{|| c}_\mathrm{c2}} = \frac{\xi_{ab}}{\xi_c}
\end{equation}
where $m_i$ ($i = ab$, $c$) are the effective mass and $\xi_i$ ($i = ab$, $c$) are GL-coherence length.
We calculate the $\xi_{i}$ by using $H^{|| ab}_\mathrm{c2}$ = $\Phi_0$/(2$\pi\xi_{ab}\xi_c$) and $H^{|| c}_\mathrm{c2}$ = $\Phi_0$/(2$\pi\xi_{ab}^2$), where $\Phi_0$ is the quantum flux.
The $\xi_{ab}$ and $\xi_{c}$ were estimated to be 24.0 and 0.742 nm.
The $\Gamma$ was obtained to be approximately 32.3, suggesting that the superconductor has remarkably high anisotropy.

As shown in Fig. \ref{fig4}(c), the thermodynamic critical field $\mu_0H_\mathrm{c}(T)$, which is directly related to the condensation energy of the superconductor through $\Delta F$ = $H_\mathrm{c}^2$/8$\pi$, was estimated by using the relationship, 
\begin{equation}
\label{ }
\Delta F(T) = F_\mathrm{n} - F_\mathrm{s} = \int_{T_\mathrm{c}}^T dT^{\prime} \int_{T_\mathrm{c}}^{T^{\prime}} \left( {C_\mathrm{n}}/{T''} - {C_\mathrm{s}}/{T''} \right) dT^{\prime\prime}.
\end{equation}
Inset of Fig. \ref{fig4}(c) shows that $H_\mathrm{c}(T)$ follows the parabolic form $H_\mathrm{c}(T)$ = $H_\mathrm{c}$(0)(1 $-$ $t^2$), where $t$ = $T/T_\mathrm{c}$, at low temperatures.
$\mu_0H_\mathrm{c}$(0) is determined unambiguously from an extrapolation to $T$ = 0, and the obtained value is $\mu_0H_\mathrm{c}$(0) = 18.5 mT.

We calculated the SC parameters for LaO$_{0.5}$F$_{0.5}$BiSSe on the basis of the $\mu_0 H_\mathrm{c2}$(0) and $\mu_0 H_\mathrm{c}$(0) values.
According to the GL theory, the GL-coherence length $\xi$(0) and GL parameter $\kappa$(0) = $\lambda$(0)/$\xi$(0) can be obtained from $\mu_0 H_\mathrm{c2}$(0), $\mu_0 H_\mathrm{c1}$(0), and $\mu_0 H_\mathrm{c}$(0) : $\mu_0 H_\mathrm{c}(0) = {\mu_0 H_\mathrm{c2}(0)}/{\sqrt{2}\kappa(0)}$, $\mu_0 H_\mathrm{c1}(0) = {\mu_0 H_\mathrm{c}(0)^2}/{H_\mathrm{c2}(0)} [\ln \kappa(0) + 0.08]$, and $\mu_0 H_\mathrm{c1}(0) = {\Phi_0}/{\pi\lambda(0)^2}$, where $\Phi_0$ is the flux quantum.
On the basis of the relationship, $\kappa$(0) was estimated as approximately 707 and 21.9. 
$\mu_0 H_\mathrm{c1}$(0) was estimated to be approximately 0.123 and 1.89 mT and $\lambda$(0) was calculated to be 2320 and 590 nm.

\section{Discussion}
Lastly, we discuss the SC gap symmetry of LaO$_{0.5}$F$_{0.5}$BiSSe.
Our measurements of $C_\mathrm{e}(T, H)$ suggest that a single SC gap is fully opened.
$\mu$SR measurements of LaO$_{0.5}$F$_{0.5}$BiS$_{2}$ show multiple SC gap.
Specific heat measurement is insensitive to light bands, because $C_\mathrm{e}(T)/T$ at low temperatures is proportional to $N(E)$.
It is likely that contribution from minor bands, which is not detected from specific heat can be seen in $\mu$SR measurements.
Theoretically, the candidates of SC gap symmetry are proposed as follows: the conventional $s$-wave, spin triplet $p$, and $d$-wave symmetries.
Although specific heat measurements imply that the superconductivity is fully gapped, $d$-wave symmetry can be realized in the disconnect Fermi surfaces.
The Fermi surface of the Se-doped compound is not elucidated at this stage.
Assuming that the Fermi surface of LaO$_{0.5}$F$_{0.5}$BiSSe contains disconnected small electron pockets at the Brillouin Zone boundary, such as the under doped Nd(O, F)BiS$_{2}$, $d_{xy}$ symmetry is ruled out and $d_{x^2 - y^2}$ symmetry can be realized.
In addition, in the case of an increase in the Se-doped value, which changes into the Fermi surface with hole pockets around the $\Gamma$ point, $s_{\pm}$ symmetry can be realized.
However, the emergence of unconventional superconductivity with sign reversal change is unlikely.
Mean free path $\ell_{ab}$ is calculated from $\ell_{ab}$ = $\mu_{0}v_\mathrm{F}^{ab}\lambda_{ab}^{2}(0)/\rho_{0}$.
Fermi velocity $v_\mathrm{F}$ is estimated to be 0.11 $\times$ 10$^{6}$ m/s by using the relationship $\xi_{ab} = 0.18(\hbar v_\mathrm{F}^{ab}$/$k_\mathrm{B}T_\mathrm{c}$).
We obtain $\ell_{ab}$ $\sim$ 10 nm, which is clearly shorter than $\xi_{ab}$.
The result suggests that superconductivity is in dirty limit.
As superconductivity with sign-reversal change is very sensitive to impurity scattering, it is likely that nodal $d$-wave state cannot be realized in the Se-doped compound.
In the case of $s_{\pm}$-wave state seen in Fe-based superconductors, the robustness against the impurities should be proved in a more complete way, because simple substitution effect could be insufficient to reveal sign-reversal change in SC gap.
Since distinct multi-gap feature can be detected from specific heat measurements of the Fe-based superconductors\cite{spe1,spe2}, the effective mass of minor bands is significantly large.
However, single-gap structure detected in the $C_\mathrm{e}(T, H)$ measurements of LaO$_{0.5}$F$_{0.5}$BiSSe indicates that the mass of the minor bands is negligible.
Therefore, we suggest that the $s$-wave structure can be the best candidate for the SC-gap symmetry of LaO$_{0.5}$F$_{0.5}$BiSSe.

\section{Conclusions}
In conclusion, to confirm bulk superconductivity and reveal the SC-gap symmetry of LaO$_{0.5}$F$_{0.5}$BiSSe, specific heat measurements were performed using a hand-made sensitive calorimeter.
Specific heat measurement shows that superconductivity of LaO$_{0.5}$F$_{0.5}$BiSSe is of bulk nature.
A large specific heat jump ($\Delta C_\mathrm{e}/\gamma T_\mathrm{c}$ = 2.31) shows strong electron--phonon coupling.
2$\Delta(0)/k_\mathrm{B}T_\mathrm{c}$ obtained from the $T$-dependence of the electronic specific heat $C_\mathrm{e}(T)$ below $T_\mathrm{c}$ is 4.5, and suggests strong electron--phonon coupling.
The value is much smaller than that obtained from STM measurements (2$\Delta/k_\mathrm{B}T_\mathrm{c}$ $\sim$ 17)\cite{delta}.
$T$- and $H$-dependences of the specific heat measurements imply that superconductivity of LaO$_{0.5}$F$_{0.5}$BiSSe is fully gapped.
These results are consistent with the $s$-wave symmetry proposed for the SC state of BiS$_2$-based superconductors.

\begin{acknowledgments}
We thank S. Kittaka for useful discussions about specific heat measurements.
This work was  supported in part by the Grant for Promotion of Niigata University Research Projects (No. 524501), the Grant-in-Aid for Young Scientists (B) (No. 26800183), the Sasakawa Scientific Research Grant (26-207) from the Japan Science Society, the Uchida Energy Science Promotion Foundation (26-1-22), and Sasaki Environment Technology Foundation.
\end{acknowledgments}

\bibliography{apssamp}

\begin{thebibliography}{9}
\bibitem{0.3} In the case of the sample with more than 1 mg (ex. LaO$_{0.5}$F$_{0.5}$Bi(S$_{0.7}$Se$_{0.3}$)$_{2}$), no upturn is seen at low temperatures in zero magnetic field.
\end{thebibliography}

\end{document}